\documentclass[11pt]{article}

\usepackage[margin=1in]{geometry}
\usepackage{setspace}
\usepackage{amsmath,amssymb,amsthm}
\usepackage{graphicx}
\usepackage{booktabs}
\usepackage{float}
\usepackage{csquotes}
\usepackage{caption}
\usepackage{subcaption}
\usepackage{threeparttable}
\usepackage{url}
\usepackage[hidelinks]{hyperref}
\usepackage[numbers,super,sort&compress]{natbib}

\begin{document}

%\doublespacing
\raggedright

\begin{titlepage}
%TC:ignore
\begin{center}
{\Large\bfseries CARB: A Covariate-Assessed Robust Borrowing Strategy with Literature-Informed Prior Weights for External Data \par}
\end{center}

\vspace{1.5em}
\textbf{Authors:}
Jinping Liang$^{1}$, Guannan Gong$^{2}$, Satrajit Roychoudhury$^{3}$, Wei Wei$^{2,*}$

\vspace{1.5em}
\textbf{Affiliations:}\\
$^{1}$ Yale School of Public Health, New Haven, Connecticut, United States \\
$^{2}$ Yale Cancer Center, Yale School of Medicine, New Haven, Connecticut, United States \\
$^{3}$ Pfizer Inc., New York, New York, United States \\

\vspace{1.5em}
\textbf{Corresponding author:} Wei Wei\\
Yale Cancer Center, Yale School of Medicine \\
New Haven, Connecticut 06520, United States\\
Email: \href{mailto:wei.wei@yale.edu}{wei.wei@yale.edu}

%TC:endignore
\end{titlepage}

\newpage
%TC:ignore
\section*{Abstract}
\textbf{Background/Aims:}
Borrowing external control data can improve the efficiency of clinical trials, particularly when patient accrual is difficult. A persistent challenge is how to prespecify the degree of borrowing systematically and transparently. In practice, prior weights are often selected heuristically or calibrated through simulation to achieve desired operating characteristics, making them difficult to justify scientifically. Moreover, patient-level covariate data from external sources are rarely available when the new trial is designed, precluding patient-level adjustment methods.

\textbf{Methods:}
We propose CARB (Covariate-Assessed Robust Borrowing), a framework that formalizes prior-weight specification as a design-stage assessment of baseline compatibility. Using only aggregate information, CARB quantifies discrepancies in prespecified baseline covariates between the new trial and each external source, without using outcome data from the new trial. A prespecified mapping translates the resulting dissimilarity measure into a source-specific prior weight on the exchangeable component of a robust borrowing model.

\textbf{Results:}
Simulation studies show that CARB reduces bias and type I error inflation relative to fixed borrowing under observed and unobserved incompatibility, while improving efficiency when external controls are compatible. An application to advanced melanoma trials illustrates covariate-informed borrowing from multiple historical sources.

\textbf{Conclusions:}
An apparent discrepancy in reported baseline characteristics serves as a warning signal that reduces borrowing. CARB provides a transparent and reproducible way to borrow cautiously when patient-level covariate data from external sources are unavailable.

\textbf{Keywords:}
hybrid control; informative prior; oncology; rare disease
%TC:endignore

\newpage
\section{Introduction}
Randomized controlled trials (RCTs) remain the gold standard for evaluating treatment effects, but adequately powered trials with large concurrent control arms can be difficult to conduct, particularly in oncology and rare-disease settings. External controls (ECs) from historical trials or real-world data may improve efficiency by reducing the required size of the concurrent control arm.\cite{ventz2022design,ventz2022use} When individual patient data (IPD) are available, weighting, matching, or outcome modeling can reduce observed covariate imbalance before Bayesian dynamic borrowing is applied.\cite{wang2019propensity,wang2022propensity,zhu2023propensity,wei2024propensity} In practice, however, external IPD are often unavailable at the design stage, leaving investigators with published outcome summaries and aggregate baseline characteristics.

Bayesian dynamic borrowing methods\cite{hobbs2011hierarchical,ibrahim2015power,schmidli2014robust,kaizer2018bayesian} can discount external information when it conflicts with the target trial. A persistent challenge is how to prespecify the initial degree of borrowing from each source. Weights are often chosen heuristically or calibrated through simulation and tipping-point analyses,\cite{best2021assessing,weru2026information} which may make their scientific rationale difficult to relate to observable characteristics of the external studies. Recent U.S. Food and Drug Administration draft guidance likewise emphasizes systematic, transparent, prespecified, and scientifically justified prior construction when external information is used.\cite{fda_bayes_draft} In particular, the guidance highlights the importance of assessing the relevance of external data, addressing prior-data conflict, and using interpretable discounting mechanisms that can be clearly documented for review. This creates a need for methods that link borrowing strength to observable information about the target trial and candidate external sources.

To address this challenge, we propose CARB (Covariate-Assessed Robust Borrowing), a framework for prespecifying source-specific borrowing weights when external IPD are unavailable. Clinical trial publications routinely report baseline characteristics in “Table 1”, including demographic variables, disease characteristics, and prognostic factors. Although these summaries do not contain sufficient information to remove confounding or support patient-level causal adjustment, they can reveal observable differences between a target trial and candidate ECs. Using summary-level covariate information, CARB formulates prior-weight specification as a baseline compatibility assessment. It quantifies discrepancies across a prespecified set of baseline covariates, summarizes them through a scalar dissimilarity measure, and maps that measure to a prior probability that each EC is exchangeable with the internal control. Greater observed dissimilarity therefore leads to less initial borrowing.

The covariates, dissimilarity measure, and mapping are specified without using target-trial outcomes. Once the required target-trial baseline summaries are available, the source-specific weights can be calculated before the outcomes are examined. At the analysis stage, Bayesian model averaging (BMA) updates support for different borrowing configurations using the observed control outcomes. CARB therefore separates covariate-informed design-stage assessment from outcome-based posterior updating. Although we implement CARB using BMA, the covariate-assessed prior weights can also be incorporated into other dynamic borrowing approaches, including meta-analytic-combined methods. \cite{neuenschwander2016use}

The proposed framework has several practical advantages. First, it replaces an otherwise heuristic prior-weight choice with a transparent rule linked to observable baseline information. Second, it accommodates multiple external sources and permits different borrowing weights according to their apparent baseline compatibility with the target trial. Third, it can be implemented using information routinely available in published clinical trial reports. Although the primary focus is on settings in which external IPD are unavailable, CARB can also be applied when IPD are available. In that setting, investigators can first use appropriate patient-level methods to improve covariate balance and then apply CARB to determine the initial degree of borrowing from the weighted or matched external data. 

The remainder of the paper is organized as follows. Section~\ref{sec:met} describes the CARB framework. Section~\ref{sec:case} illustrates the approach through a case study involving multiple external controls. Section~\ref{sec:simu} evaluates operating characteristics of CARB under varying degrees of covariate and outcome discordance. Section~\ref{sec:diss} concludes with a
discussion of limitations, extensions, and future directions.

\section{Methods}\label{sec:met}
\subsection{Framework and compatibility assessment}\label{sec:framework}
Consider a target RCT with an internal experimental arm (IE), an internal control arm (IC), and $J$ candidate external controls $EC_1,\ldots,EC_J$. We assume that candidate ECs use the same control or standard of care and share a prespecified set of key baseline covariates with the target trial. CARB uses these covariates to assess source-level compatibility and prespecify the initial degree of borrowing from each EC.

The primary setting considered here is one in which external IPD are unavailable and only published aggregate baseline and outcome summaries can be used. Candidate ECs should therefore be selected based on substantive comparability in key prognostic factors and treatment-effect modifiers, control intervention, eligibility criteria and outcome ascertainment, while following systematic evidence-review principles to reduce selection bias. \cite{pocock1976combination, lim2018minimizing} Borrowing then relies on a study-level exchangeability assumption: conditional on the reported baseline summaries and study characteristics, the mean control outcome in $EC_j$ is assumed exchangeable with that in the IC. This assumption is inherently more uncertain than conditional exchangeability with IPD, because it cannot be strengthened through patient-level causal preprocessing. As a result, borrowing from such sources should generally be more conservative. 

CARB addresses this challenge by using published baseline summaries to quantify cross-trial dissimilarity and then translating that dissimilarity into a source-specific prior exchangeability probability. It does not attempt to remove confounding through causal adjustment, but instead uses observable study-level information to assess compatibility and borrows information in a transparent and prespecified manner.

When IPD are available, weighting or matching may first be used to improve observed covariate balance before CARB is applied. We describe how CARB can be applied in the IPD setting in Supplementary Appendix A.1.

\subsection{Assessing cross-trial dissimilarity using baseline covariates}\label{sec:dis}
For an EC without IPD, let $\bar{x}^{EC_j}_k$ and $v^{EC_j}_k$ denote the published mean (or proportion) and variance for covariate $X_k$, and let $\bar{x}^{R}_k$ and $v^{R}_k$ denote the corresponding quantities in the target RCT. To quantify imbalance between the target RCT and external control $EC_j$, we use the standardized mean difference (SMD).\cite{austin2009balance} For covariate $X_k$, the SMD between $EC_j$ and the target trial is
\begin{equation}\label{eq:sdiff}
d_{kj}
=
\frac{
\left|\bar{x}^{EC_j}_k-\bar{x}^R_k\right|
}{
\sqrt{\tfrac{1}{2}\left(v^{EC_j}_k+v^R_k\right)}
}.
\end{equation}

The SMD is scale invariant and widely used to assess covariate balance in matching and weighting studies. Smaller values of $d_{kj}$ indicate closer alignment in the observed distribution of covariate $X_k$ between the two studies.

To summarize overall dissimilarity between $EC_j$ and the target RCT, we define
\begin{equation}\label{eq:dissim}
D_j=\max_{k=1,\ldots,p} d_{kj}, \qquad j=1,\ldots,J.
\end{equation}

Using the maximum SMD yields a conservative summary that prioritizes protection against severe imbalance in any single key prognostic covariate. Alternative aggregation strategies, including weighted averages reflecting prognostic importance, are discussed in Section~\ref{sec:diss}.

\subsection{Mapping dissimilarity to prior exchangeability}\label{sec:map}
For the binary endpoint considered in the main formulation, let $\theta_{IC}$ denote the response probability in the IC arm and let $\theta_j$ denote the response probability in external control $EC_j$. For each $EC_j$, we introduce an exchangeability indicator $Z_j\in\{0,1\}$, where $Z_j=1$ indicates that $EC_j$ is treated as exchangeable with the IC at the source level, such that $\theta_j=\theta_{IC}$. Thus, exchangeability here refers to equality of the marginal response probabilities across sources, rather than patient-level conditional exchangeability.

To encode uncertainty on the exchangeability assumption, CARB assigns a prior probability of exchangeability as a smooth monotone decreasing function of the dissimilarity measure:
\begin{equation}\label{eq:omega_def}
\Pr(Z_j=1\mid D_j)
=
\pi_j(D_j)
=
\frac{1}{1+\exp(a_j+b_jD_j)},
\end{equation}
where $a_j\in\mathbb{R}$ and $b_j>0$ are prespecified design parameters. 

The quantity $\pi_j(0)$ represents the prior plausibility of exchangeability under ideal observed balance; it need not equal 1 because unmeasured prognostic factors, differences in trial conduct, endpoint assessment, or temporal changes may still induce non-exchangeability. The value of $\pi_j(0)$ should be elicited from experts at the design stage. Given $\pi_j(0)$, 
\begin{equation}\label{eq:aj_def}
a_j = \log\!\left(\frac{1-\pi_j(0)}{\pi_j(0)}\right).
\end{equation}

The parameter $b_j$ controls how rapidly exchangeability decreases with covariate imbalance. Investigators specify a reference dissimilarity $D^*$ at which exchangeability is negligible, $\pi_j(D^*)=\epsilon$, where $\epsilon$ is small (e.g., 0.01). Given prespecified $a_j$, $D^*$ and $\epsilon$, we have
\begin{equation}\label{eq:bj_def}
b_j = \frac{\log\!\left(\frac{1-\epsilon}{\epsilon}\right)-a_j}{D^*}.
\end{equation}

\begin{figure}[htbp]
    \centering
\includegraphics[width=0.6\textwidth]{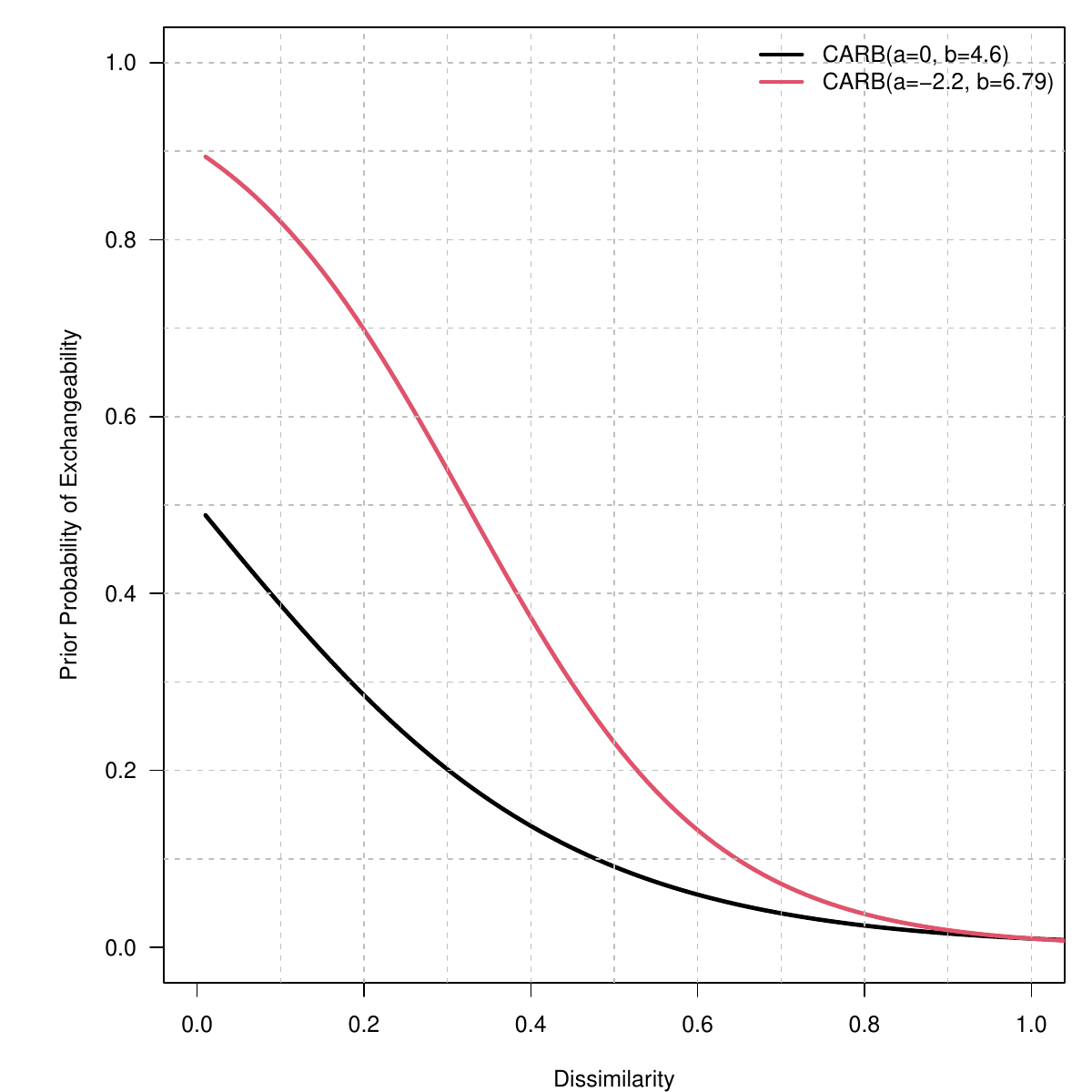}
    \caption{Mapping baseline covariate dissimilarity to prior exchangeability probability. The horizontal axis shows the dissimilarity measure $D_j$, defined as the maximum standardized mean difference across shared baseline covariates between the target trial and an external control. The vertical axis shows the corresponding prior exchangeability probability $\pi_j(D_j)$ between an external control and the internal control of the target trial, conditional on observed differences in baseline covariates.  In the advanced melanoma case study, we prespecify $D^*=1$, $w_0=\pi_j(0)=0.5$, and $\epsilon=0.01$ at the design stage, which yields the curve $a_j=0$ and $b_j=4.60$. }
    \label{fig:dis}
\end{figure}

Thus, the CARB calibration is completely determined by three user-specified quantities: $\pi_j(0)$, $D^*$, and $\epsilon$. These quantities can be prespecified at the design stage using only baseline covariate considerations and expert judgment. No outcome data from the target RCT are used. Figure~\ref{fig:dis} illustrates the mapping from $D_j$ to the prior exchangeability probability $\pi_j(D_j)$. For example, setting $\pi_j(0)=0.5$, $\epsilon = 0.01$ and $D^*=1$ yields $a_j=0, b_j=4.6$. This mapping process produces a smooth, monotone decreasing curve that starts at 0.5 when $D_j=0$ (perfect covariate balance) and reduces to 0.01 when $D_j=D^*=1$, thereby enforcing increasingly conservative borrowing as cross-trial dissimilarity grows. 

%CARB is not more complicated than the common practice of fixing a prior weight at 0.5 or tuning it by simulation. Rather, it provides a systematic, transparent, and less biased way to specify that prior weight by linking it directly to observed covariates imbalance. It also allows $\pi_j(0)$ to differ across external controls, which is useful when expert confidence differs by source. This source-specific formulation provides finer control over prior uncertainty than a single global borrowing weight.

%A practical implication is that ECs with IPD may support more confident borrowing than ECs without IPD. When IPD is available, causal preprocessing methods such as weighting or matching can be used to align the observed covariate distribution of the EC with that of the target trial. To the extent that this preprocessing improves covariate balance, it strengthens the assumption of source-level exchangeability and builds a stronger case for borrowing. By contrast, when IPD is unavailable, causal preprocessing cannot be performed and compatibility must be judged solely from published aggregate summaries. As a result, any observed covariate differences remain uncorrected and may appear larger than they would after weighting or matching, leading to greater uncertainty about exchangeability and more cautious borrowing. Importantly, this does not imply that ECs with IPD are intrinsically more relevant; rather, IPD allows their comparability to the target trial to be assessed and improved more directly with respect to observed covariates.

\subsection{Bayesian model averaging over borrowing configurations}\label{sec:bma}
With $J$ external controls, there are $K=2^J$ possible borrowing configurations. Let $\mathcal{M}_k$ denote the model associated with configuration
\[
\mathbf{Z}^{(k)}=\bigl(Z_1^{(k)},\ldots,Z_J^{(k)}\bigr),
\]
where $Z_j^{(k)}\in\{0,1\}$ indicates whether $EC_j$ is treated as exchangeable under model $\mathcal{M}_k$. That is, we have $\theta_j=\theta_{IC}$ under 
$\mathcal{M}_k$ if $Z_j^{(k)}=1$.

For each EC, let $r_{EC_j}$ and $n_{EC_j}$ denote its effective number of responses and effective sample size. For ECs with IPD, these quantities are obtained from the weighted data; for ECs without IPD, the reported response rate is preserved while the effective sample size is capped at the IC sample size to prevent a large external source from dominating the analysis. Details are provided in Supplementary Appendix A.2.

Under configuration $\mathcal{M}_k$, external controls classified as exchangeable contribute to an informative Beta prior for the IC response probability:
$$
\alpha_k = \alpha_0+\sum_{j=1}^{J} Z_j^{(k)}r_{EC_j}, \qquad
\beta_k = \beta_0+\sum_{j=1}^{J}Z_j^{(k)} (n_{EC_j}-r_{EC_j}),
$$
where $\alpha_0$ and $\beta_0$ are small positive constants
(e.g., $\alpha_0=\beta_0=1$).

Given $r_{IC}$ responses among $n_{IC}$ patients, the configuration-specific posterior is
\[
\theta_{IC}\mid r_{IC},n_{IC},\mathcal{M}_k
\sim
\mathrm{Beta}
\left(
\alpha_k+r_{IC},
\beta_k+n_{IC}-r_{IC}
\right).
\]

Assuming source-specific exchangeability indicators are independent a priori, the prior probability of configuration $\mathcal{M}_k$ is
\begin{equation}
\label{eq:prior_config}
\Pr(\mathcal{M}_k)
=
\prod_{j=1}^{J}
\pi_j(D_j)^{Z_j^{(k)}}
\{1-\pi_j(D_j)\}^{1-Z_j^{(k)}}.
\end{equation}

After observing the IC outcomes, these configuration probabilities are updated by Bayes' rule using their marginal likelihoods. The resulting CARB posterior averages over uncertainty in the borrowing configuration:
\begin{equation}
\label{eq:post}
\pi(\theta_{IC}\mid\mathrm{data})
=
\sum_{k=1}^{K}
\pi(\theta_{IC}\mid\mathrm{data},\mathcal{M}_k)
\Pr(\mathcal{M}_k\mid\mathrm{data}).
\end{equation}

Thus, baseline compatibility determines the prior support for each borrowing configuration through $\pi_j(D_j)$, while the observed IC outcomes subsequently update the relative support for those configurations. The marginal likelihood calculation and additional implementation details are provided in Supplementary Appendix A.3.1. BMA for continuous outcomes is described in Supplementary Appendix A.3.2.

\subsection{Posterior inference for treatment effect}\label{sec:posterior}
For binary efficacy endpoints, interest typically lies in testing
\[
H_0:\theta_{IE}\le \theta_{IC}
\qquad \text{versus} \qquad
H_1:\theta_{IE}>\theta_{IC},
\]
where $\theta_{IE}$ is the parameter of interest for the outcomes in the IE arm of the target trial. 

Let $r_{IE}$ denote the number of responses among the $n_{IE}$ patients in the experimental arm. We assume $r_{IE}\sim \text{Binomial}(n_{IE},\theta_{IE}), \quad \theta_{IE}\sim \text{Beta}(\alpha_0,\beta_0)$, which yields the posterior
\[
\pi(\theta_{IE}\mid r_{IE},n_{IE})
=
\text{Beta}\bigl(\alpha_0+r_{IE},\;\beta_0+n_{IE}-r_{IE}\bigr).
\]

%Let $\mathbf{r}=(r_{EC_1},\ldots,r_{EC_J},r_{IC},r_{IE})$ and $\mathbf{n}=(n_{EC_1},\ldots,n_{EC_J},n_{IC},n_{IE})$. 
Assume $\theta_{IC}$ has a posterior distribution defined by Equation \ref{eq:post}. We declare the experimental treatment effective if
\begin{equation*}\label{eq:decision_rule}
\Pr\!\left(
\theta_{IE}>\theta_{IC}
\;\middle|\;
r_{IC}, n_{IC}, r_{IE}, n_{IE}, \mathbf{r}_{EC}, \mathbf{n}_{EC}\right)>\phi,
\end{equation*}
where $\phi\in(0,1)$ is a prespecified decision threshold, for example $\phi=0.95$.

\section{A case study in advanced melanoma}\label{sec:case}
\subsection{Case study setting and data sources}
To illustrate CARB, we considered several pembrolizumab trials in advanced melanoma. We designated the Lambrolizumab trial \cite{hamid2013safety} as the target trial, with pembrolizumab 10 mg/kg every 3 weeks (Q3W) as the internal control (IC) arm and 10 mg/kg every 2 weeks (Q2W) as the internal experimental (IE) arm. Two published external controls (ECs) were included: the pembrolizumab 10 mg/kg Q3W arms from KEYNOTE-002\cite{ribas2015pembrolizumab} and KEYNOTE-006\cite{schachter2017pembrolizumab}. Because IPD were unavailable for these trials, CARB used published baseline summaries and overall response rates. We additionally included a hypothetical external control (HEC) with simulated IPD that was propensity-score weighted to the target trial population to illustrate the use of CARB when IPD are available.

Supplementary Table 1 summarizes the detailed trial information of each EC and the target trial. Table~\ref{tab:case_covariates} presents the shared baseline characteristics and the corresponding standardized mean differences (SMDs) comparing each EC with the IC arm of the target trial. After weighting, the HEC shows the closest alignment with the target trial, with a maximum SMD of 0.08. KEYNOTE-002 and KEYNOTE-006 exhibit larger observed baseline covariate differences, with maximum SMDs of 0.32 and 0.61, respectively. The largest discrepancies involved ECOG performance status in KEYNOTE-002 and BRAF mutation status and prior chemotherapy in KEYNOTE-006. 

For comparison, we considered a uniform-prior analysis, CARB, and a covariate-assessed meta-analytic-combined approach (CA-MAC) based on the MAC framework.\cite{neuenschwander2016use} In CA-MAC, the source-specific exchangeability probability is determined by the CARB mapping, $Z_j\sim\mathrm{Bernoulli}\{\pi_j(D_j)\}$, rather than by a fixed prior probability. Full model and computational details for MAC and CA-MAC are provided in Supplementary Appendix A.4.

\subsection{Results}
For the case study, we prespecified $D^*=1$, $\pi_j(0)=0.5$, and $\epsilon=0.01$, yielding $a_j=0$ and $b_j=4.60$. Applying this mapping to the observed maximum SMDs gave prior exchangeability probabilities of 0.410 for the HEC, 0.185 for KEYNOTE-002, and 0.057 for KEYNOTE-006.

Table~\ref{tab:priorconfig} shows the different CARB borrowing configurations involving the HEC, KEYNOTE-002, and KEYNOTE-006, together with their prior and posterior probabilities. Configurations containing KEYNOTE-006 received little prior and posterior weight, whereas those involving the HEC and KEYNOTE-002 received greater support because of their closer baseline and outcome agreement with the IC.

\begin{figure}[H]
    \centering
    \includegraphics[width=0.8\textwidth]{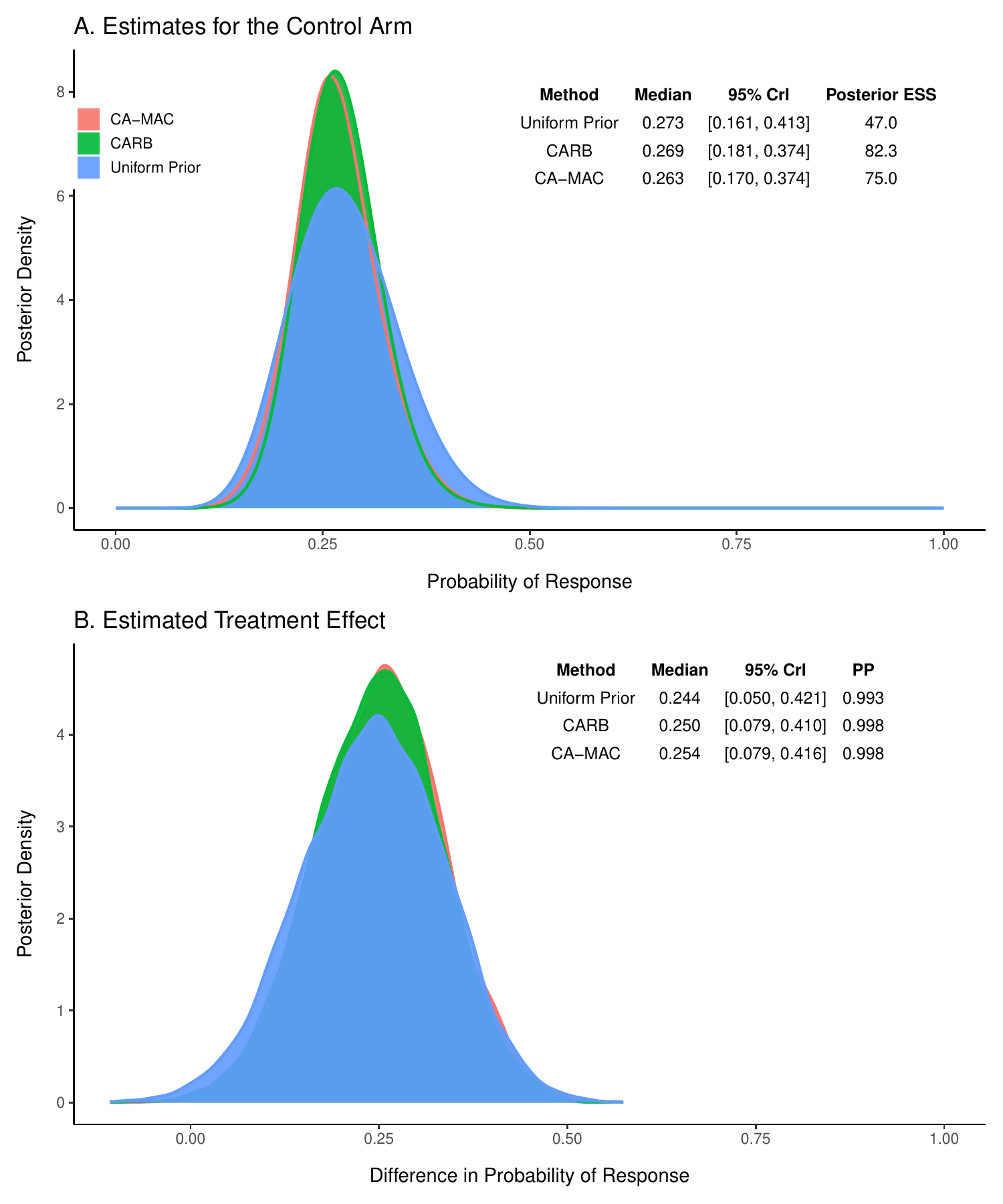}
    \caption{Posterior inference for the control arm and treatment effect in the advanced melanoma case study under three approaches: no borrowing with uniform priors, CARB, and the covariate-assessed meta-analytic-combined method (CA-MAC). Panel A shows the posterior distributions for the response probability in the internal control (IC) arm. Panel B shows the posterior distributions for the treatment effect. Both borrowing approaches yield narrower posterior intervals than the no-borrowing analysis, while preserving evidence of a favorable treatment effect.}
    \label{fig:case}
\end{figure}

Figure~\ref{fig:case} compares the posterior distributions for the IC response rate and the treatment effect under the three methods. Posterior median IC response rates were similar, but CARB and CA-MAC provided greater precision than the uniform-prior analysis. Compared to the analysis based on uniform prior, CARB increased the posterior effective sample size (ESS) from 47.0 to 82.3 and narrowed the 95\% credible interval for the IC response rate from [0.161, 0.413] to [0.181, 0.374]; CA-MAC yielded a posterior ESS of 75.0 and a 95\% credible interval of [0.170, 0.374]. We calculated the posterior ESS for CARB and CA-MAC using the expected local information ratio method. \cite{neuenschwander2020predictively} For the treatment effect, the posterior median difference in response probability is similar across methods. CARB and CA-MAC also produce narrower credible intervals as a result of borrowing from the most compatible ECs while limiting the impact of less comparable sources.

Overall, the case study demonstrates how CARB uses observable baseline compatibility to determine source-specific borrowing while limiting the influence of less comparable external controls.

\section{Simulation studies}\label{sec:simu}
\subsection{Design rationale and simulation settings}
We conducted simulation studies to evaluate the operating characteristics of CARB under varying degrees of covariate and outcome compatibility between the RCT and an external control (EC). 

Suppose there is an RCT with $n_R=90$ patients randomized 2:1 to the internal experimental (IE) and internal control (IC) arms, together with one EC of $n_{EC}=200$ patients. The primary endpoint was response rate. Only aggregate baseline covariate and outcome summaries from the EC were used by CARB. Five independent baseline covariates were generated, with $X_1$ continuous and $X_2,\ldots,X_5$ binary. Outcomes were generated from
\[
\mathrm{logit}(\theta)
=
\gamma_{0s}+A_i\Delta+\sum_{k=1}^{5}X_k\gamma_k,
\]

where $A_i$ is the treatment indicator, $\Delta$ is the treatment effect on the log-odds scale, and $\gamma_{0s}$ is a study-specific intercept for $s\in\{\mathrm{RCT},\mathrm{EC}\}$.

We considered four scenarios representing different sources of compatibility or incompatibility between the EC and IC:
\begin{itemize}
    \item Full compatibility in observed covariates and outcomes, the ideal case for borrowing
    \item Imbalance in prognostic covariate $X_5$, representing observed confounding
    \item Imbalance in non-prognostic covariate $X_1$ ($\gamma_1=0$), to evaluate if CARB can avoid over-penalizing imbalance in covariates that do not affect the endpoint
    \item Unmeasured confounding, similar observed covariates but different study-specific intercepts ($\gamma_{0,\text{RCT}}\neq\gamma_{0,\text{EC}}$) and outcome distributions. 
\end{itemize}

\begin{figure}[H]
    \centering
    \includegraphics[width=0.8\textwidth]{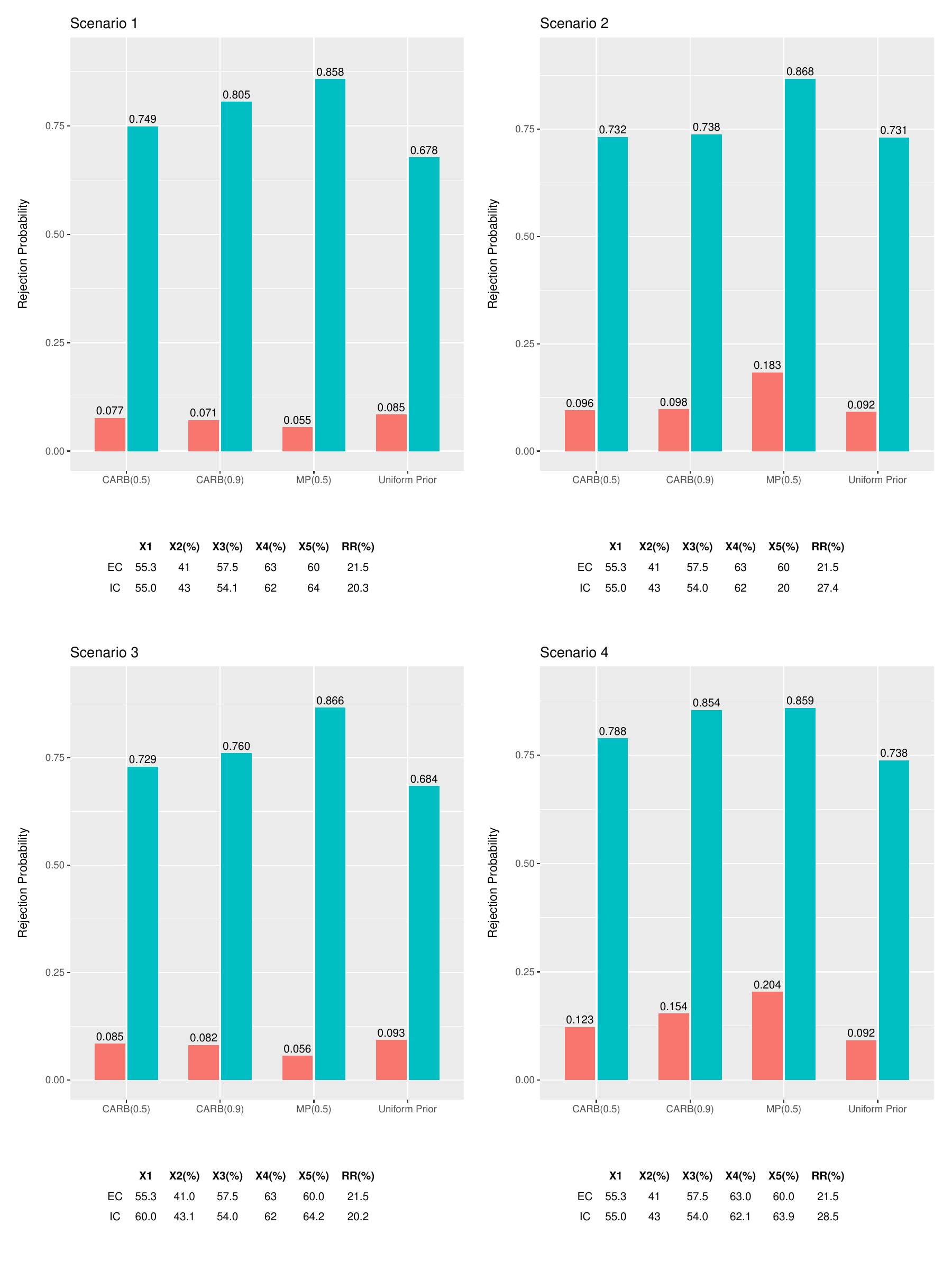}
    \caption{Operating characteristics under varying compatibility scenarios. Type I error rate (red) and power (blue) across simulation scenarios representing full compatibility (Scenario 1), prognostic imbalance (Scenario 2), non-prognostic imbalance (Scenario 3), and outcome discordance due to unobserved confounder (Scenario 4).}
    \label{fig:opers}
\end{figure}

Tables inserted in Figure \ref{fig:opers} summarize the simulated covariate profiles and response rates for the EC and IC under each scenario. For each scenario, the EC was fixed and 10,000 RCTs were simulated under the null ($\Delta=0$) and alternative ($\Delta=1$).

We compared the following approaches in modeling the response rate for the IC arm:
\begin{itemize}
    \item Uniform prior (no borrowing)
    \item A two-component mixture prior with a fixed borrowing weight of 0.5(Mixture Prior (0.5)), consisting of a vague $Beta(1,1)$ component and an informative $Beta(1 + r_{EC},\, 1 + n_{EC}-r_{EC})$ component constructed from the EC
    \item CARB with $\pi_j(0)=0.5, D^*=1, \epsilon = 0.01$, yielding
    $a=0$ and $b=4.60$
    \item CARB with $\pi_j(0)=0.9, D^*=1, \epsilon = 0.01$, yielding $a=-2.2$ and $b=6.79$
\end{itemize}
The second setup represents a commonly used borrowing strategy in which the prior mixing weight is fixed and does not depend on baseline covariate compatibility. A uniform prior was used for the IE response rate in all analyses. Across 10,000 simulations per scenario under the null and the alternative, we evaluated type I error rate, power, and bias in the estimated treatment effect.

\subsection{Results}
Figure~\ref{fig:opers} summarizes rejection probabilities across the four scenarios, and Table~\ref{tab:bias_iqr} reports the median and interquartile range of treatment-effect bias. Overall, CARB balanced robustness and efficiency. Across scenarios, CARB generally provides better control of type I error rate than the fixed mixture prior (MP), while preserving much of the efficiency gain from borrowing when the EC is compatible with the target trial. The more aggressive CARB calibration with $\pi_j(0)=0.9$ tends to borrow more heavily and therefore achieves slightly higher power than $\pi_j(0)=0.5$, but at the cost of some increase in type I error rate when incompatibility is present. 

Under full compatibility, all borrowing methods improve efficiency relative to the analysis based on a uniform prior. Type I error rates were reasonably controlled under the null for all methods. Under the alternative, borrowing leads to clear gains in power, with rejection probabilities of 0.749 and 0.805 for the two CARB calibrations, compared with 0.858 for MP and 0.678 for the uniform prior. The treatment-effect bias remained near zero for all methods. Thus, when the EC is truly compatible, CARB preserves most of the efficiency gain from borrowing without introducing meaningful bias.

When the EC differed from the IC in a prognostic covariate, the fixed MP was anti-conservative, with a type I error rate of 0.183. In contrast, CARB yielded type I error rates of 0.096 and 0.098, close to the uniform-prior value of 0.092. The bias table clarifies why this occurs: under both $H_0$ and $H_1$, MP shows substantial positive median bias, whereas both CARB versions remain centered much closer to zero. This scenario illustrates the main benefit of CARB: when observed incompatibility is driven by an imbalance in a prognostic covariate, CARB appropriately discounts the EC and protects against bias and type I error rate inflation.

When imbalance occurred only in a non-prognostic covariate, CARB remains appropriately robust without over-penalizing borrowing. Under the null, rejection probabilities are 0.085 and 0.082 for CARB, compared with 0.056 for Mixture Prior(0.5) and 0.093 for the uniform prior. Under the alternative, CARB yields rejection probabilities of 0.729 and 0.760, intermediate between Mixture Prior(0.5) (0.866) and the uniform prior (0.684). Because CARB discounts sources based on observed covariate discrepancy regardless of whether that covariate is prognostic, this scenario highlights the importance of selecting clinically relevant covariates for the compatibility assessment.

Finally, when the observed covariates were similar but the outcome distributions differed because of unmeasured confounding, all borrowing methods showed some type I error inflation. CARB remained more conservative than the fixed mixture prior, with type I error rates of 0.123 and 0.154 versus 0.204. The bias table shows the same ordering: MP has the largest positive bias, and CARB with $\pi_j(0)=0.5$ is more conservative. Thus, although CARB cannot completely eliminate the impact of unmeasured confounding, posterior updating provides additional protection relative to fixed borrowing.

\section{Discussion}\label{sec:diss}
By grounding prior borrowing weights in observed baseline covariate information and providing full prespecification at the design stage, CARB provides a more interpretable and regulator-aligned approach to borrowing external data in clinical trials.
Compared with the common practice of using simulation to select a prior weight for the informative component, CARB is straightforward to implement. It requires only a small number of interpretable design inputs, including the prior exchangeability probability under ideal covariate balance and a reference dissimilarity threshold ($D^*$) beyond which borrowing is negligible ($\epsilon$). These quantities can be fully prespecified at the design stage using expert judgment and baseline covariate information alone, without using outcome data from the target trial.

An important feature of CARB is the intentionally asymmetric interpretation of baseline information. A substantial discrepancy in a prespecified, clinically important baseline characteristic serves as a warning signal and supports reducing the initial degree of borrowing. Conversely, close agreement among the reported baseline characteristics does not establish that the target trial and external source are exchangeable. Differences may remain in unreported covariates, joint covariate distributions, outcome assessment, trial conduct, or other study features. Thus, CARB uses baseline summaries to identify observable reasons for caution rather than to demonstrate causal comparability. This conservative approach has recognized limitations. CARB may discount an external source when the observed baseline discrepancy is unrelated to the outcome, thereby sacrificing potential efficiency. Conversely, it cannot detect incompatibility arising solely from unmeasured or unreported factors when the reported baseline summaries appear similar. These two limitations are illustrated by the simulation scenarios involving imbalance in a non-prognostic covariate and outcome discordance due to an unmeasured factor, respectively. They reflect the restricted information available without IPD and should be considered when selecting baseline covariates, calibrating the dissimilarity-to-weight mapping, and conducting sensitivity analyses.

An important consideration in any borrowing framework is control of the type I error rate. Recent theoretical work has shown that strict control of the type I error under all possible data-generating mechanisms may preclude meaningful power gains from incorporating external information, even when dynamic borrowing is used. \cite{kopp2020power} This result highlights an inherent trade-off between robustness and efficiency. Rather than enforcing uniform type I error control under every conceivable scenario, CARB targets strong protection under clinically plausible forms of incompatibility while preserving the possibility of efficiency gains when external data are sufficiently aligned with the target trial. This perspective is consistent with regulatory guidance, which emphasizes transparency, prespecification, and scientific justification, rather than overly conservative calibration that may nullify the practical value of borrowing.

In this work, we used the maximum standardized difference across covariates to define the scalar dissimilarity measure. This choice is conservative and easy to interpret, but it may overemphasize a single covariate difference that is not strongly prognostic for the endpoint. One possible extension is to construct a weighted dissimilarity measure, where the weights reflect the prognostic importance of individual covariates. Such an approach could allow CARB to distinguish more clearly between imbalance in highly prognostic covariates and imbalance in covariates with limited clinical relevance. However, specification of these weights may be challenging, particularly when IPD is unavailable.

By leveraging literature-extracted baseline covariates, CARB provides a scalable bridge between the expanding corpus of published clinical trials and Bayesian evidence synthesis. Published trial reports almost always include structured summaries of baseline characteristics in Table 1, providing information on key demographic and prognostic variables. These summaries therefore represent a rich but underused source of evidence about cross-trial similarity. Recent advances in AI-based tools specialized for processing clinical trial reports, such as Lead-Onc\cite{gong2026leadonc}, further increase the feasibility of extracting and standardizing such information at scale. In this way, CARB offers a practical framework for turning routinely reported baseline summaries into principled borrowing decisions.

\newpage
%TC:ignore
\section*{Declaration of conflicting interests}
The authors declare that there is no conflict of interest.

\section*{Funding}
This work was supported by the National Cancer Institute (P30-CA16359) and the National Institute of Dental and Craniofacial Research (P50DE030707).

\section*{Data availability statement}
The data and R code used to reproduce the case study analysis are publicly available on GitHub at \url{https://github.com/allisoncjj/CARB.git}.
%TC:endignore

\newpage
\bibliographystyle{SageV}
\bibliography{references}

\newpage

\newpage

\newpage

\newpage
\begin{table}[H]
\centering
\caption{Baseline Characteristics and Standardized Mean Differences (SMDs) between the internal control (IC) and each external control (EC). For the HEC, baseline summaries are shown after propensity score weighting. For KEYNOTE-002 and KEYNOTE-006, baseline summaries are taken directly from published aggregate reports, as patient-level data are unavailable. The SMD for each covariate is computed relative to the IC arm of the target trial. The maximum SMD within each EC is shown in bold and is used by CARB as the  dissimilarity measure for mapping baseline imbalance to a prior exchangeability probability.
}
\label{tab:case_covariates}
\begin{tabular}{l c cc cc cc}
\toprule
 & IC 
 & \multicolumn{2}{c}{HEC} 
 & \multicolumn{2}{c}{KEYNOTE002} 
 & \multicolumn{2}{c}{KEYNOTE006} \\
\cmidrule(lr){3-4} 
\cmidrule(lr){5-6} 
\cmidrule(lr){7-8}
Variable 
 &  
 & Mean & SMD 
 & Mean & SMD 
 & Mean & SMD \\
\midrule
Age, Mean(SD)        & 61.0 & 60.5 & 0.03 & 60.0 & 0.07 & 63.0 & 0.13 \\
Female (\%)   & 41.1 & 45.0 & 0.08& 39.8 & 0.03 & 37.2 & 0.08 \\
ECOG=0 (\%)      & 69.6 & 70.0 & 0.01 & 54.1 & \textbf{0.32} & 68.2 & 0.03 \\
BRAF mutant (\%) & 12.5 & 10.0 & 0.08 & 22.1 & 0.26 & 35.3 & 0.55 \\
Normal LDH (\%) & 71.7 & 73.0 & 0.03 & 59.0 & 0.27 & 64.1 & 0.16 \\
Prior Chemotherapy (\%) & 41.1 & 45.0 & \textbf{0.08} & 46.4 & 0.11 & 14.8 & \textbf{0.61} \\
\bottomrule
\end{tabular}
\end{table}

\newpage
\begin{table}[H]
\centering
\caption{Prior and posterior probabilities of different borrowing configurations in the advanced melanoma case study. Each row represents a possible borrowing configuration formed by including or excluding the three external controls: HEC, KEYNOTE-002 (KN002), and KEYNOTE-006 (KN006). An indicator value of 1 denotes inclusion of the corresponding external control in the informative prior for the IC response rate, whereas 0 denotes exclusion. The parameters $\alpha$ and $\beta$ are the resulting Beta prior parameters under each configuration. The “Prior” column reports the covariate-assessed configuration probabilities, and the “Posterior” column reports the posterior configuration probabilities after incorporating the observed IC data.}\label{tab:priorconfig}
\begin{tabular}{cccccccc}
  \hline
 Model& HEC & KN002 & KN006 & $\alpha$ & $\beta$ & Prior Prob & Posterior Prob \\ 
  \hline
1 & 0 & 0 & 0 & 1 & 1 & 0.454 & 0.159 \\ 
  2 & 1 & 0 & 0 & 13 & 35 & 0.315 & 0.487 \\ 
  3 & 0 & 1 & 0 & 12 & 35 & 0.103 & 0.157 \\ 
  4 & 1 & 1 & 0 & 25 & 69 & 0.071 & 0.126 \\ 
  5 & 0 & 0 & 1 & 17 & 30 & 0.027 & 0.025 \\ 
  6 & 1 & 0 & 1 & 29 & 64 & 0.019 & 0.029 \\ 
  7 & 0 & 1 & 1 & 29 & 63 & 0.006 & 0.009 \\ 
  8 & 1 & 1 & 1 & 41 & 98 & 0.004 & 0.008 \\ 
   \hline
\end{tabular}
\end{table}

\newpage

\begin{table}[H]
\centering
\caption{Median and interquartile range of bias in the estimated treatment effect under the null hypothesis ($H_0$) and the alternative hypothesis ($H_1$) across the four simulation scenarios. Scenario 1: fully compatible EC and IC. Scenario 2: imbalance in a prognostic covariate. Scenario 3: imbalance in a non-prognostic covariate. Scenario 4: outcome discordance due to unmeasured confounding.}
\label{tab:bias_iqr}
\begin{tabular}{l c ccc ccc}
\toprule
& & \multicolumn{3}{c}{$H_0$} & \multicolumn{3}{c}{$H_1$} \\
\cmidrule(lr){3-5} \cmidrule(lr){6-8}
Method & Scenario & 50\% & 25\% & 75\% & 50\% & 25\% & 75\% \\
\midrule
CARB($w_0=0.5$) & 1 & -0.002 & -0.051 & 0.047 & -0.006 & -0.063 & 0.050 \\ 
CARB($w_0=0.9$) & 1 & 0.001 & -0.046 & 0.046 & -0.004 & -0.055 & 0.048 \\ 
Mixture Prior(0.5) & 1 & -0.006 & -0.046 & 0.035 & -0.009 & -0.056 & 0.039 \\ 
Uniform Prior & 1 & -0.006 & -0.068 & 0.045 & -0.011 & -0.074 & 0.052 \\ 
\hline
CARB($w_0=0.5$) & 2 & -0.002 & -0.066 & 0.061 & -0.004 & -0.077 & 0.061 \\ 
CARB($w_0=0.9$) & 2 & -0.002 & -0.066 & 0.061 & -0.002 & -0.077 & 0.061 \\ 
Mixture Prior(0.5) & 2 & 0.043 & -0.011 & 0.083 & 0.033 & -0.017 & 0.082 \\ 
Uniform Prior & 2 & -0.004 & -0.068 & 0.059 & -0.004 & -0.080 & 0.061 \\ 
\hline
CARB($w_0=0.5$) & 3 & -0.002 & -0.055 & 0.052 & -0.009 & -0.066 & 0.049 \\ 
CARB($w_0=0.9$)& 3 & 0.000 & -0.049 & 0.050 & -0.007 & -0.061 & 0.049 \\ 
Mixture Prior(0.5) & 3 & -0.004 & -0.038 & 0.035 & -0.008 & -0.055 & 0.039 \\ 
Uniform Prior & 3 & -0.006 & -0.067 & 0.056 & -0.012 & -0.075 & 0.051 \\ 
\hline
CARB($w_0=0.5$) & 4 & 0.015 & -0.047 & 0.072 & 0.009 & -0.056 & 0.068 \\ 
CARB($w_0=0.9$) & 4 & 0.029 & -0.025 & 0.081 & 0.023 & -0.035 & 0.079 \\ 
Mixture Prior(0.5) & 4 & 0.045 & -0.006 & 0.092 & 0.039 & -0.017 & 0.088 \\ 
Uniform Prior & 4 & -0.004 & -0.068 & 0.060 & -0.004 & -0.080 & 0.060 \\ 
\bottomrule
\end{tabular}
\end{table}

\end{document}

%\doublespacing
\raggedright

\begin{center}
{\Large\bfseries CARB Supplementary Appendix \par}
\end{center}

\section*{A.1 Causal identification and preprocessing when IPD are available}
Consider a target RCT with an internal experimental arm (IE) and an internal control arm (IC). For subject $i$, let $A_i \in \{0,1\}$ denote treatment assignment, where $A_i=1$ indicates assignment to the IE and $A_i=0$ indicates assignment to the IC. Let $\mathbf{X}_i=(X_{i1},\ldots,X_{ip})$ denote a vector of baseline covariates, and let $Y_i$ denote the observed outcome.

Suppose a set of $J$ external control (EC) sources is identified, denoted $EC_1,\ldots,EC_J$. Let $S_i \in \{RCT,EC_1,\ldots,EC_J\}$ denote the data-source indicator. We assume that the same control or standard of care (SOC) is used in the ECs and in the IC arm of the target RCT. We further assume that a set of $p$ key baseline covariates is available in the target RCT and in each EC.

Let $Y_i(0)$ and $Y_i(1)$ denote the potential outcomes under the control and experimental treatment, respectively. Valid integration of ECs requires standard identifying assumptions including (1) consistency of the observed outcome under the treatment received and the corresponding potential outcome; (2) positivity of trial participation, which requires sufficient overlap in baseline covariates between the target RCT and each EC; and (3) the conditional exchangeability of potential outcomes across data sources given measured baseline covariates ($Y(0)\perp S \mid \mathbf{X}$).

These assumptions are rarely verifiable in practice. Even after careful study selection and adjustment for observed covariates, residual non-exchangeability may remain due to unmeasured prognostic factors, differences in trial conduct, endpoint assessment, or temporal changes in standard of care. As a result, an effective borrowing strategy should not only account for observed covariate imbalance, but also encode expert knowledge on the relevance and credibility of each candidate EC in prior formulation. This motivates CARB: first reduce observed imbalance when possible, then represent the remaining uncertainty about source-level exchangeability within a Bayesian model.

For ECs with individual patient data (IPD), CARB begins with a causal preprocessing step that aims to align the observed covariate distribution of the ECs with that of the target RCT.

Let
\[
e(\mathbf{X}_i)=\Pr(S_i=RCT\mid \mathbf{X}_i)
\]
denote the propensity score of belonging to the target RCT. The propensity score model is fit using patients from the IC arm of the target RCT together with patients from ECs for which IPD is available.

The propensity score can be estimated using a logistic regression model. Alternatively, the covariate balancing propensity scores (CBPS), estimated by incorporating covariate balance into the optimization procedure, can be used. The CBPS is shown to be more robust to model misspecification than standard logistic-regression-based propensity scores.\cite{imai2014covariate}

Using the estimated propensity score, we construct a weighted pseudo-population by assigning each control subject weight
\[
w_i=
\begin{cases}
1, & S_i=RCT,\\[0.4em]
\dfrac{e(\mathbf{X}_i)}{1-e(\mathbf{X}_i)}, & S_i\neq RCT.
\end{cases}
\]

For external control $EC_j$, let
\[
\bar{x}^{EC_j}_k
=
\frac{\sum_{i\in EC_j} w_i x_{ik}}{\sum_{i\in EC_j} w_i},
\qquad k=1,\ldots,p,
\]
denote the weighted sample mean or weighted proportion for covariate $X_k$. 

For a continuous covariate, we define the weighted variance as
\[
v^{EC_j}_k
=
\frac{\sum_{i\in EC_j} w_i (x_{ik}-\bar{x}^{EC_j}_k)^2}{\sum_{i\in EC_j} w_i}.
\]

For a binary covariate, we use
\[
v^{EC_j}_k
=
\bar{x}^{EC_j}_k\bigl(1-\bar{x}^{EC_j}_k\bigr).
\]

Let
\[
\theta_j
=
\mathbb{E}_w\{Y(0)\mid S=EC_j\}
=
\mathbb{E}_w(Y\mid A=0,S=EC_j)
\]
denote the weighted mean outcome for external control $EC_j$, and let
\[
\theta_{IC}
=
\mathbb{E}(Y\mid A=0,S=RCT)
\]
denote the mean outcome in the internal control arm.

Under consistency, conditional exchangeability, and positivity, causal preprocessing aims to make weighted control outcomes from $EC_j$ exchangeable to those in the target RCT, so that
\begin{equation*}\label{eq:mean_exchangeability}
\theta_j=\theta_{IC}.
\end{equation*}
Thus, this step seeks to translate conditional exchangeability at the individual level into marginal exchangeability of control outcomes at the source level.

Although we focus on weighting in this work, CARB is not tied to a single causal adjustment strategy. Matching can be incorporated within the same workflow.

\section*{A.2 Effective sample size construction}
Let $r_{IC}$ denote the number of responses in the internal control arm, with sample size $n_{IC}$. For an external control with IPD, define the weighted number of responses and weighted sample size as
\[
r_{EC_j}=\sum_{i\in EC_j} w_i y_i,
\qquad
n_{EC_j}=\sum_{i\in EC_j} w_i,
\]
where $y_i=1$ indicates a favorable response. 

By weighting each $EC_j$ to the IC population, the expected weighted sample size of an individual EC satisfies
\begin{equation}\label{eq:ewn}
    \mathbb{E}\!\left(\sum_{i\in S_i} w_i\right)=n_{IC},
\end{equation}
thereby reducing the risk that posterior inference is dominated by a large external dataset. We provide proof for \ref{eq:ewn} at the end of this section.

For an EC without IPD, let $n^*_{EC_j}$ and 
$r^*_{EC_j}$ denote the sample size and number of responses reported in literature, we consider
$$r_{EC_j} = r^*_{EC_j} \times \min\left(1, \frac{n_{IC}}{n^*_{EC_j}}\right),
\qquad
n_{EC_j} = \min\left(n^*_{EC_j}, n_{IC}\right).
$$  
Likewise, the observed response rate is preserved while the effective sample size is capped at $n_{IC}$. This ensures that no single summary-data EC can dominate posterior inference purely because of its original sample size.

The following proposition formalizes a useful property of odds weighting when an external control is reweighted to the target-trial population. In particular, it shows that the expected weighted sample size of the external control equals the target-trial sample size. Under a just-identified covariate balancing propensity score (CBPS) fit with an intercept included among the balancing functions, the corresponding sample equality holds exactly.

\begin{proposition}
Let $S_i\in\{0,1\}$ denote the source indicator, where $S_i=1$ if subject $i$ belongs to the target trial and $S_i=0$ if subject $i$ belongs to the external control. Let
\[
e(X_i)=\Pr(S_i=1\mid X_i),
\qquad
w_i=\frac{e(X_i)}{1-e(X_i)}.
\]
Then the expected weighted sample size of the external control satisfies
\[
\mathbb{E}\!\left(\sum_{i:S_i=0} w_i\right)
=
\mathbb{E}\!\left(\sum_{i=1}^N S_i\right).
\]
In particular, if the target-trial sample size is fixed at $n_R$, then
\[
\mathbb{E}\!\left(\sum_{i:S_i=0} w_i\right)=n_R.
\]

Moreover, under a just-identified CBPS fit for the target population with an intercept included among the balancing functions, the sample analogue satisfies
\[
\sum_{i:S_i=0}\frac{\hat e(X_i)}{1-\hat e(X_i)}
=
\sum_{i=1}^N S_i
=
n_R
\]
exactly, provided the balancing equations are solved exactly.
\end{proposition}

\begin{proof}
The weighted external-control sample size can be written as
\[
\sum_{i:S_i=0} w_i
=
\sum_{i=1}^N (1-S_i)\frac{e(X_i)}{1-e(X_i)}.
\]
Taking expectation and applying iterated expectation,
\[
\mathbb{E}\!\left[\sum_{i=1}^N (1-S_i)\frac{e(X_i)}{1-e(X_i)}\right]
=
\sum_{i=1}^N
\mathbb{E}\!\left[
\mathbb{E}\!\left\{(1-S_i)\frac{e(X_i)}{1-e(X_i)}\mid X_i\right\}
\right].
\]
Since $e(X_i)$ is fixed given $X_i$,
\[
\mathbb{E}\!\left\{(1-S_i)\frac{e(X_i)}{1-e(X_i)}\mid X_i\right\}
=
\frac{e(X_i)}{1-e(X_i)}\,\mathbb{E}(1-S_i\mid X_i).
\]
But
\[
\mathbb{E}(1-S_i\mid X_i)=1-e(X_i),
\]
so
\[
\mathbb{E}\!\left\{(1-S_i)\frac{e(X_i)}{1-e(X_i)}\mid X_i\right\}=e(X_i).
\]
Therefore,
\[
\mathbb{E}\!\left(\sum_{i:S_i=0} w_i\right)
=
\sum_{i=1}^N \mathbb{E}\{e(X_i)\}
=
\sum_{i=1}^N \mathbb{E}(S_i)
=
\mathbb{E}\!\left(\sum_{i=1}^N S_i\right).
\]
If the target-trial sample size is fixed by design at $n_R$, this reduces to
\[
\mathbb{E}\!\left(\sum_{i:S_i=0} w_i\right)=n_R.
\]

For the CBPS result, note that the just-identified ATT-style balancing equation with intercept $\tilde X_i\equiv 1$ implies
\[
\sum_{i=1}^N
\left[
S_i-\frac{\hat e(X_i)(1-S_i)}{1-\hat e(X_i)}
\right]
=0.
\]
Rearranging gives
\[
\sum_{i:S_i=0}\frac{\hat e(X_i)}{1-\hat e(X_i)}
=
\sum_{i=1}^N S_i.
\]
Hence, when the target-trial size is fixed at $n_R$,
\[
\sum_{i:S_i=0}\frac{\hat e(X_i)}{1-\hat e(X_i)}=n_R.
\]
This completes the proof.
\end{proof}

\section*{A.3 Bayesian model averaging in CARB}
\subsection*{A.3.1 Binary outcomes}
Under model $\mathcal{M}_k$, we define an informative $Beta(\alpha_k, \beta_k)$ prior for the probability of response (i.e, response rate) in the IC arm (i.e., $\theta_{IC}$) by pooling outcome data from the external controls treated as exchangeable:
$$
\alpha_k = \alpha_0 + \sum_{j=1}^J Z_j^{(k)} r_{EC_j}, \qquad
\beta_k = \beta_0 + \sum_{j=1}^J Z_j^{(k)} (n_{EC_j}-r_{EC_j}),
$$
where $\alpha_0$ and $\beta_0$ are small positive constants, for example $\alpha_0=\beta_0=1$.

Assume $r_{IC}\mid \theta_{IC}\sim \text{Binomial}(n_{IC},\theta_{IC})$, the posterior distribution of $\theta_{IC}$ under model $\mathcal{M}_k$ is
\[
\pi(\theta_{IC}\mid r_{IC},n_{IC},\mathcal{M}_k)
=
\text{Beta}\bigl(\alpha_k+r_{IC},\;\beta_k+n_{IC}-r_{IC}\bigr).
\]

Assuming the exchangeability indicators are independent a priori, the prior probability of model $\mathcal{M}_k$ is
\begin{equation*}\label{eq:prior_config}
\Pr(\mathcal{M}_k)
=
\prod_{j=1}^{J}
\pi_j(D_j)^{Z_j^{(k)}}
\bigl\{1-\pi_j(D_j)\bigr\}^{1-Z_j^{(k)}}.
\end{equation*}

The marginal likelihood of the IC data under model $\mathcal{M}_k$ is
\begin{equation*}\label{eq:marginal_config}
\Pr(r_{IC}\mid \mathcal{M}_k)
\propto
\frac{
\mathrm{B}\bigl(\alpha_k+r_{IC},\;\beta_k+n_{IC}-r_{IC}\bigr)
}{
\mathrm{B}(\alpha_k,\beta_k)
},
\end{equation*}
where $\mathrm{B}(\cdot,\cdot)$ denotes the Beta function.

The posterior probability of model $\mathcal{M}_k$ is 
\begin{equation*}\label{eq:post_model_prob}
\Pr(\mathcal{M}_k\mid r_{IC})
=
\frac{
\Pr(r_{IC}\mid \mathcal{M}_k)\Pr(\mathcal{M}_k)
}{
\sum_{s=1}^{K}
\Pr(r_{IC}\mid \mathcal{M}_s)\Pr(\mathcal{M}_s)
}.
\end{equation*}

Let $\mathbf{r}_{EC}=(r_{EC_1}, \dots, r_{EC_J})$ and
$\mathbf{n}_{EC}=(n_{EC_1}, \dots, n_{EC_J})$. The CARB posterior for $\theta_{IC}$ is then obtained by Bayesian model averaging:
\begin{equation*}\label{eq:post}
\pi(\theta_{IC}\mid r_{IC},n_{IC}, \mathbf{r}_{EC}, \mathbf{n}_{EC})
=
\sum_{k=1}^{K}
\pi(\theta_{IC}\mid r_{IC},n_{IC},\mathcal{M}_k)
\Pr(\mathcal{M}_k\mid r_{IC}).
\end{equation*}

\subsection*{A.3.2 Continuous outcomes}
For continuous endpoints, we assume the outcomes in the internal control (IC) arm follow a normal distribution with an unknown mean $\theta_{IC}$ and unknown variance $\sigma^2$. Let $Y_{IC} = \{y_{1}, \dots, y_{n_{IC}}\}$ denote the observed outcomes in the IC arm, with sample mean $\bar{y}_{IC}$ and sample variance $v_{IC} = \frac{1}{n_{IC}}\sum_{i=1}^{n_{IC}} (y_i - \bar{y}_{IC})^2$.

For an external control with individual patient data (IPD), define the weighted sample size, weighted mean, and weighted variance as:$$n_{EC_j} = \sum_{i \in EC_j} w_i, \quad \bar{y}_{EC_j} = \frac{\sum_{i \in EC_j} w_i y_i}{n_{EC_j}}, \quad v_{EC_j} = \frac{\sum_{i \in EC_j} w_i (y_i - \bar{y}_{EC_j})^2}{n_{EC_j}}. $$ If IPD are unavailable, these aggregate summaries can be extracted directly from published reports, setting $w_i = 1$ up to the capped sample size $n_{IC}$, analogous to the binary case. 

\vspace{0.5em}
\noindent
\textbf{Prior distribution under model $\mathcal{M}_k$} \\
To model both the unknown mean and variance, we utilize a Normal-Inverse-Gamma (NIG) conjugate prior. We begin with a weakly informative base prior defined by hyperparameters $(\mu_0, n_0, \alpha_0, \beta_0)$. Under model $\mathcal{M}_k$, the indicator $Z_j^{(k)} \in \{0,1\}$ determines whether external control $EC_j$ is treated as exchangeable. By pooling the summary statistics of the exchangeable external controls, the informative prior for model $\mathcal{M}_k$ is updated as: 
$$\theta_{IC}, \sigma^2 | \mathcal{M}_k \sim \text{NIG}(\mu_k, n_k, \alpha_k, \beta_k). $$

The pooled hyperparameters are defined as: 
$$n_k = n_0 + \sum_{j=1}^J Z_j^{(k)} n_{EC_j}$$
$$\mu_k = \frac{n_0 \mu_0 + \sum_{j=1}^J Z_j^{(k)} n_{EC_j} \bar{y}_{EC_j}}{n_k}$$
$$\alpha_k = \alpha_0 + \frac{1}{2} \sum_{j=1}^J Z_j^{(k)} n_{EC_j}$$
$$\beta_k = \beta_0 + \frac{1}{2} \sum_{j=1}^J Z_j^{(k)} n_{EC_j} v_{EC_j} + \frac{1}{2} \left[ n_0 \mu_0^2 + \sum_{j=1}^J Z_j^{(k)} n_{EC_j} \bar{y}_{EC_j}^2 - n_k \mu_k^2 \right]$$
Here, $\beta_k$ incorporates both the weighted within-trial variances and the between-trial variance of the exchangeable sources.

\vspace{0.5em}
\noindent
\textbf{Posterior distribution under model $\mathcal{M}_k$} \\
Given the observed outcomes $Y_{IC}$ in the target trial, the posterior distribution under model $\mathcal{M}_k$ updates analytically to another Normal-Inverse-Gamma distribution:$$\theta_{IC}, \sigma^2 | Y_{IC}, \mathcal{M}_k \sim \text{NIG}(\mu_k^*, n_k^*, \alpha_k^*, \beta_k^*).$$
The posterior hyperparameters are:
$$n_k^* = n_k + n_{IC}$$
$$\mu_k^* = \frac{n_k \mu_k + n_{IC} \bar{y}_{IC}}{n_k^*}$$
$$\alpha_k^* = \alpha_k + \frac{n_{IC}}{2}$$
$$\beta_k^* = \beta_k + \frac{n_{IC}}{2} v_{IC} + \frac{n_k n_{IC}}{2 n_k^*} (\bar{y}_{IC} - \mu_k)^2$$

\vspace{0.5em}
\noindent
\textbf{Marginal likelihood and model averaging} \\\
The marginal likelihood of the IC data under model $\mathcal{M}_k$ obtained by integrating out $\theta_{IC}$ and $\sigma^2$, has a closed-form analytic solution given by:
$$Pr(Y_{IC} | \mathcal{M}_k) = (2\pi)^{-\frac{n_{IC}}{2}} \sqrt{\frac{n_k}{n_k^*}} \frac{\Gamma(\alpha_k^*)}{\Gamma(\alpha_k)} \frac{\beta_k^{\alpha_k}}{(\beta_k^*)^{\alpha_k^*}}. $$

The posterior probability of model $\mathcal{M}_k$ is then calculated exactly as in the binary case: $$Pr(\mathcal{M}_k | Y_{IC}) = \frac{Pr(Y_{IC} | \mathcal{M}_k) Pr(\mathcal{M}_k)}{\sum_{s=1}^K Pr(Y_{IC} | \mathcal{M}_s) Pr(\mathcal{M}_s)}. $$

The final CARB posterior for the continuous treatment effect is obtained by averaging the $\mathcal{M}_k$-specific marginal posteriors of $\theta_{IC}$ using these updated configuration weights $Pr(\mathcal{M}_k | Y_{IC})$.

\section*{A.4 Implementation of the covariate-assessed MAC approach}
For the MAC approach, we assume the response rate from 
each control arm is
$$
\theta_j = Z_j\theta_{j0} + (1-Z_j)\theta_{j1},
$$
for $j\in\{EC_1, \dots, EC_J, IC\}$. 

We estimate $\theta_{j0}$ by sharing information across different control arms
$$logit(\theta_{j0})\sim Normal(\mu, \sigma^2),$$
whereas $\theta_{j1}$ is estimated independently without borrowing information from other arms
$$logit(\theta_{j1})\sim Normal(0, 4).$$

For $j\in\{EC_1, \dots, EC_J\}$, the indicator $Z_j$ determines if an EC contributes to the estimation of $\theta_{IC}$ and follows a Bernoulli prior 
$$Z_j\sim Bernoulli(p_j).$$ 
Unlike the original MAC approach which assumes a fixed value for $p_j$, we follow the CARB principle and consider
$$Z_j\sim Bernoulli(\pi_j(D_j)),$$ 
where $\pi_j(D_j)$ is determined through the mapping procedure described in Section 2.3. 

The IC is always included in the exchangeable component, $Z_{IC}=1$. %For the IC, we assume $Z_j=1$.

We consider a normal prior for mean $\mu$
$$\mu\sim Normal(0, 4)$$
and a truncated normal prior for standard deviation $\sigma$
$$\sigma \sim Normal(0, 0.25)T(0, ).$$

Using \textit{JAGS}, we implemented the MAC method by taking 50000 posterior samples and discarding the first 30000 samples. We refer to this extension as the covariate-assessed meta-analytic-combined approach (CA-MAC).

\newpage 
\bibliographystyle{SageV}
\bibliography{references}

\newpage
\begin{center}
{\Large\bfseries CARB Supplementary Tables \par}
\end{center}

\begin{table}[htbp]
\centering
\caption{Trials included in the case study. The Lambrolizumab trial is designated as the
target trial for illustration, with the pembrolizumab 10 mg/kg Q3W arm treated as the
internal control (IC) and the pembrolizumab 10 mg/kg Q2W arm treated as the internal
experimental (IE) arm. The pembrolizumab 10 mg/kg Q3W arms from KEYNOTE-002 and
KEYNOTE-006 are treated as published external controls (ECs) without individual patient
data, whereas the simulated hypothetical external control (HEC) is included to represent
an EC with individual patient data available. For each arm, the table reports the number
of responses, sample size, and observed response rate. For the HEC, these numbers are
obtained after propensity score weighting.}
\label{tab:case}
\resizebox{\textwidth}{!}{%
\begin{tabular}{lcccccc}
\hline
& Trial & Dose \& Schedule of Pembrolizumab & Arm &
Number of Responses & Sample Size & Response Rate \\
\hline
1 & HEC after weighting & 10 mg/kg Q3W & EC & 12.1 & 46.5 & 0.26 \\
2 & KEYNOTE-002 & 10 mg/kg Q3W & EC & 46 & 181 & 0.25 \\
3 & KEYNOTE-006 & 10 mg/kg Q3W & EC & 100 & 277 & 0.36 \\
4 & Lambrolizumab & 10 mg/kg Q3W & IC & 12 & 45 & 0.27 \\
5 & Lambrolizumab & 10 mg/kg Q2W & IE & 27 & 52 & 0.52 \\
\hline
\end{tabular}%
}
\end{table}